\documentclass[11pt,a4paper]{article}
\usepackage{jcappub}

\usepackage{physics}
\usepackage{mathtools}

\usepackage{tikz}
\definecolor{sym_vertex}{rgb}{.7,.7,.7}

\usepackage{bm}
\usetikzlibrary{arrows.meta}
\usetikzlibrary{calc}
\newcommand{\identity}{\text{\usefont{U}{bbold}{m}{n}1}}
\MakeRobust{\identity}

\renewcommand*{\vec}[1]{\bm{#1}}
\def\[#1\]{\begin{equation}\begin{aligned}[b]#1\end{aligned}\end{equation}} % numbering \[\] at last line

\title{Phase-space perturbation theory for cosmic large-scale structure}
\author{Hannes Heisler\footnote{\label{fa} HH and MS have contributed equally to this work.},}
\emailAdd{heisler@thphys.uni-heidelberg.de}
\author{Marvin Sipp\footref{fa}}
\emailAdd{sipp@thphys.uni-heidelberg.de}
\author{and Matthias Bartelmann}
\emailAdd{bartelmann@uni-heidelberg.de}
\affiliation{Institut für Theoretische Physik, Universität Heidelberg, Philosophenweg 12, 69120 Heidelberg, Germany}

\arxivnumber{2512.07396}

\newcommand{\dirac}{\delta_\mathrm{D}}
\newcommand{\Om}{\ensuremath\Omega_\mathrm{m}}

\newcommand{\ini}{{(\mathrm{i})}}

\newcommand{\im}{\mathrm{i}}
\newcommand{\e}{\mathrm{e}}
\newcommand{\gqp}{g_{0}^{xp}}
\newcommand{\gpp}{g_{0}^{pp}}

\newcommand{\C}{\mathcal{C}}

\abstract{We consider a perturbative approach to the Vlasov--Poisson system for cosmic structure formation that does not rely on any truncation of the momentum-cumulant hierarchy. The generally non-trivial linear solution is computed by solving a Volterra-type integral equation and higher orders are obtained recursively. As expected, the results of Eulerian standard perturbation theory are recovered for perfectly cold initial conditions. Deviating slightly from the latter by introducing a homogeneous and isotropic initial velocity dispersion, we show that all higher momentum cumulants are generated dynamically at any perturbative order. We support our numerical solutions by an analytical large-scale approximation. Our approach serves as a basis for exploring different background--perturbation splits of the phase-space density and nonperturbative techniques.}

\begin{document}
    \maketitle
    
    \section{Introduction}
    The cosmological concordance model, while tremendously successful at explaining a wealth of observed phenomena, relies on ingredients whose fundamental nature is not well understood. In particular, most of the Universe's matter content seems to interact only gravitationally and is therefore referred to as \emph{dark matter}. Driven by their gravitational instability, small fluctuations in the early dark matter distribution are enhanced, giving rise to the cosmic large-scale structure observed today. The latter serves as one of the key observables in cosmology, as its statistical properties encode information about both dark matter and the underlying laws of gravity on cosmological scales. Given that cosmic structure formation is a highly non-linear process, it is intrinsically challenging to model. Numerical simulations have proven to be very successful at reproducing the non-linear clustering~\cite[e.\,g.][]{Springel2005,Klypin2011,Springel2017,Wang2020}. Yet, their substantial computational demands make it impractical to explore a wide range of cosmological models or theories of dark matter and gravity. To address this limitation and gain deeper physical insight, analytical approaches are desirable.
    
    On scales much larger than the typical inter-particle separation but well below the cosmic horizon, the dynamics of the cold dark matter (CDM) distribution is described by the Vlasov equation (sometimes called collisionless Boltzmann equation),
    which is a conservation equation for the phase-space density~\cite{Peebles1980}. Instead of working in full phase space, momentum cumulants can be projected out, leading to an infinite hierarchy of equations in configuration space. Ignoring velocity dispersion and higher moments, this hierarchy can be self-consistently truncated. In this so-called single-stream approximation (SSA), CDM is modelled as a perfect fluid governed by the Euler--Poisson system of equations. This is the basis of Eulerian standard perturbation theory (SPT), where the density and velocity fields are expanded perturbatively around their mean values~\cite{Bernardeau2002}. While perturbation theory fails on small scales, where the density contrast grows to order unity, it proves to be a valid description on the largest scales. On intermediate scales, where perturbative approaches are still expected to work but non-linearities become important, however, the SSA misses quantitatively important physical effects~\cite{Pueblas2009}. Efforts to overcome these shortcomings include higher order truncations~\cite{Buchert1997,McDonald2011,Erschfeld2019,Erschfeld2024,Garny2023lin,Garny2023nonlin,Garny2025a,Garny2025} or absorbing small-scale physics into operators of an effective field theory~\cite{Baumann2012,Carrasco2012}. Lagrangian perturbation theory, on the other hand, models the dynamics of cold dark matter in terms of a displacement field relative to initial positions of fluid elements~\cite{Buchert1992,Buchert1993,Buchert1994,Porto2014,Rampf2021}. In order to correctly model the dynamics, though, one has to keep track of individual stream crossings, where the mapping of the initial to the evolved density contrast becomes singular~\cite{Rampf2021}. A related approach based on microscopic particle dynamics in phase space is kinetic field theory (KFT)~\cite{Bartelmann2016,Lilow2019}. In the limit of large number densities, the Hamiltonian particle dynamics are equivalent to that of the Vlasov equation~\cite{Chavanis2008}.
    Alternatively, perturbative expansions can be built directly based on the latter~\cite{Valageas2001,Nascimento2025}. However, due to the self-consistency of the SSA, they exactly reproduce SPT for perfectly cold initial conditions (CICs), as does KFT~\cite{Sipp2025}.
    
    In this work, we consider a perturbative approach to the Vlasov equation that does not rely on any truncation of its momentum-cumulant hierarchy and straightforwardly generalizes to arbitrary initial conditions and different background--perturbation splits of the phase-space density. In section~\ref{sec:kinetic_theory}, we briefly review the Vlasov--Poisson system, its moment hierarchy and SPT. Thereafter, we develop the perturbative expansion in section~\ref{sec:perturbation_theory} and show in section~\ref{sec:perfectly_cold} that for CICs we indeed recover SPT. As an example for going beyond the SSA, we include a small initial velocity dispersion in section~\ref{sec:slightly_warm}, demonstrating that all higher momentum cumulants are generated dynamically at any order in perturbation theory. We present our conclusions and give an outlook to future work in section~\ref{sec:conclusions}.
    
    \section{The Vlasov equation and its moment hierarchy}\label{sec:kinetic_theory}
    In terms of conformal time~$\tau$, comoving coordinates~$\vec{x}$ and the canonical momentum for particles of mass $m$ on a perturbed Friedmann-Lemaître-Robertson-Walker background with scale factor~$a$, $\vec{p} = a m\dv{\vec{x}}{\tau}$, the Vlasov equation reads
    \[
        \dv{f}{\tau} = \pdv{f}{\tau} + \frac{\vec{p}}{m a}\cdot\nabla_x f - am\nabla_x \Phi\cdot\nabla_p f = 0,
        \label{eq:vlasov}
    \]
    where $f$ denotes the phase-space density.
    The Newtonian gravitational potential $\Phi$ satisfies a Poisson equation,
    \[
        \Delta\Phi(\vec{x},\tau) = \frac{3}{2}\,\Omega_\mathrm{m}(\tau)\,\mathcal{H}^2(\tau)\,\delta(\vec{x},\tau),
    \]
    with the matter density parameter~$\Omega_\mathrm{m}$, the conformal Hubble function~$\mathcal{H} = \partial_\tau a / a$ and the density contrast~$\delta = (\rho-\bar{\rho})/\bar{\rho}$.
    Defining momentum moments of the phase-space density,
    \[
        \rho(\vec{x},\tau) &= \int\dd[3]{p}\, f(\vec{x},\vec{p},\tau), \\  \vec{\Pi}(\vec{x},\tau) &= \int\dd[3]{p}\, \vec{p}\, f(\vec{x},\vec{p},\tau), \\  T(\vec{x},\tau) &= \int\dd[3]{p} \vec{p} \otimes \vec{p}\, f(\vec{x},\vec{p},\tau),\ \dots
    \]
    the Vlasov equation can be rewritten as an infinite hierarchy of equations in configuration space~\cite{Peebles1980}. Decomposing the momentum moments into a bulk and a residual cumulant part,
    \[
        \vec{\Pi} = \rho \, a m \, \vec{u}, \quad T = a^2 m^2 \rho\,\qty(\vec{u} \otimes \vec{u} + \sigma),\quad\dots,
    \]
    the hierarchy can be truncated self-consistently by setting the velocity dispersion $\sigma$ to zero. This is the well-known single-stream approximation (SSA) modelling the dark matter as a perfect fluid described by an Euler--Poisson system of equations \cite{Bernardeau2002}. The SSA provides an accurate description of the CDM dynamics on large scales, where the density contrast is small, $\abs{\delta} \ll 1$, and the peculiar velocity is much smaller than the Hubble drift, $\abs{\nabla\cdot\vec{u}} \ll \mathcal{H}$. Perturbation theory can be applied in this regime. In the linear approximation of the Euler--Poisson equations, all Fourier modes decouple,
    \[
        \delta^{(1)}(\vec{k},\tau) = D(\tau)\,\delta^\ini(\vec{k}),
    \]
    and the dynamics of the so-called growth factor $D$ are described by the linear growth equation
    \[
        \dv[2]{D}{\tau} + \mathcal{H}\dv{D}{\tau} - \frac{3}{2}\,\Omega_\mathrm{m}\,\mathcal{H}^2 D = 0.
    \] 
    The growth equation has two solutions, a growing mode $D_+$ and a decaying mode $D_-$. In the limit of late times, the decaying mode can be neglected. In order to go beyond linear theory, the Euler--Poisson system can be expanded in its non-linearities, leading to the well known Eulerian standard perturbation theory (SPT), see ref.~\cite{Bernardeau2002} for a comprehensive review.
    Higher order corrections to the density contrast $\delta$ and the velocity divergence\footnote{Vorticity modes can be shown to decay in SPT and are thus typically neglected~\cite{Bernardeau2002}.} $\theta = \nabla\cdot\vec{u}$ are usually expressed in terms of the (symmetric) integration kernels
    \[
        \delta^{(n)}(\vec{k},\tau) &= D_+^n(\tau) \int \qty(\prod_{i=1}^{n} \frac{\dd[3]{q_i}}{(2\pi)^3} \, \delta^\ini(\vec{q}_i)) \,(2\pi)^3\,\dirac \qty(\vec{k}-\sum_{i=1}^{n} \vec{q}_i) F_n(\vec{q}_1,\ldots,\vec{q}_{n}), \\
        \frac{\theta^{(n)}(\vec{k},\tau)}{\mathcal{H}(\tau) f_+(\tau)} &= -D_+^n(\tau) \int \qty(\prod_{i=1}^{n} \frac{\dd[3]{q_i}}{(2\pi)^3} \, \delta^\ini(\vec{q}_i)) \,(2\pi)^3\,\dirac \qty(\vec{k}-\sum_{i=1}^{n} \vec{q}_i) G_n(\vec{q}_1,\ldots,\vec{q}_{n}),
    \]
    where $f_+ = \dv{\log D_+}{\log a}$.
    Here, only the fastest growing modes $\propto D_+^n$ are kept, while subdominant contributions are neglected. The integration kernels obey the recursion relation
    \[
        F_n(\vec{q}_1,\dots,\vec{q}_n) &= \sum_{m=1}^{n-1} \frac{G_m(\vec{q}_1,\dots,\vec{q}_m)}{(2n+3)(n-1)} 
        \begin{aligned}[t]
            \big[&(2n+1)\alpha(\vec{q}_{1\dots m}, \vec{q}_{m+1\dots n}) F_{n-m}(\vec{q}_{m+1},\ldots,\vec{q}_{n}) 
        \\ &+ 2 \beta(\vec{q}_{1\dots m}, \vec{q}_{m+1\dots n}) G_{n-m}(\vec{q}_{m+1},\ldots,\vec{q}_{n})\big],
        \end{aligned}\\
        G_n(\vec{q}_1,\dots,\vec{q}_n) &= \sum_{m=1}^{n-1} \frac{G_m(\vec{q}_1,\dots,\vec{q}_m)}{(2n+3)(n-1)} 
        \begin{aligned}[t]
            \big[&3\alpha(\vec{q}_{1\dots m}, \vec{q}_{m+1\dots n}) F_{n-m}(\vec{q}_{m+1},\ldots,\vec{q}_{n}) 
        \\ &+ 2 n \beta(\vec{q}_{1\dots m}, \vec{q}_{m+1\dots n}) G_{n-m}(\vec{q}_{m+1},\ldots,\vec{q}_{n})\big],
        \label{eq:recursion_spt}
        \end{aligned}
    \]
    where we introduced the functions
    \[
        \alpha(\vec{q}_1,\vec{q}_2) = \frac{(\vec{q}_1+\vec{q}_2)\cdot \vec{q}_1}{q_1^2}, \quad \beta(\vec{q}_1,\vec{q}_2) = \frac{(\vec{q}_1+\vec{q}_2)^2(\vec{q}_1\cdot \vec{q}_2)}{2 q_1^2 q_2^2},
    \]
    and the notation $\vec{q}_{1\dots n} = \vec{q}_1+ \dots+\vec{q}_n$. The above expansion is typically derived for an Einstein-de Sitter (EdS) cosmology, but it has been shown to be an excellent approximation even in the presence of a cosmological constant, as $\Om/f_+^2 \approx 1$ during most of the late-time evolution \cite{Scoccimarro1998}.
    From the recursion \eqref{eq:recursion_spt}, perturbative corrections to any order can be obtained straightforwardly. While this works well on large scales, perturbation theory breaks down in the strongly non-linear regime.
    Furthermore, as SPT is based on the SSA, shell-crossing effects are not captured. On intermediate scales, where perturbative techniques can still be applied, this leads to different predictions compared to the full phase-space dynamics and misses physical effects like the dynamical generation of vorticity \cite{Pueblas2009}.
    This motivates the construction of a perturbative expansion of the Vlasov equation not relying on any cumulant truncation, which is presented in the following section.

    \section{Perturbation theory of the Vlasov equation}\label{sec:perturbation_theory}

    We introduce the time coordinate~$\eta = \log( \frac{D(\tau)}{D^\ini})$, which is a common and convenient choice for cosmic structure formation~\cite{Nusser1998}. In terms of the (non-canonical) momentum~$\vec{p} = m \dv{\vec{x}}{\eta}$, the Newtonian equations of motion can be formulated as
    \[
        \dv{\vec{p}}{\eta} = -\qty(\frac{3}{2}\frac{\Omega_\mathrm{m}}{f_+^2}-1) \, \vec{p} - m \nabla_x \Tilde{\Phi} \approx -\frac{\vec{p}}{2} - m \nabla_x \Tilde{\Phi} \qq{with} \Delta_x \Tilde{\Phi} =  \frac{3}{2}\frac{\Omega_\mathrm{m}}{f_+^2} \delta \approx \frac{3}{2} \delta,
        \label{eq: particle_dynamics}
    \]
    where $\tilde{\Phi}= \Phi/ (\mathcal{H}f_+)^2$ is a rescaled potential and the EdS approximation was applied in the last step.
    In Fourier space, the Vlasov equation then reads
    \[\label{eq:Vlasov_Fourier}
       \qty[\partial_\eta - \vec{k}\cdot \nabla_l + \frac{1}{2} \, \vec{l}\cdot \nabla_l] f(\vec{k},\vec{l},\eta) - \frac{3}{2 \bar{\rho}} \int \frac{\dd[3]{k'}}{(2\pi)^3} \, \frac{\vec{k}'\cdot \vec{l}}{k'^2} f\qty(\vec{k}-\vec{k}',\vec{l},\eta) \, f\qty(\vec{k}',0,\eta) = 0,
    \]
    where $(\vec{k},\vec{l})$ are the Fourier conjugates to $(\vec{x},\vec{p}/m)$.
    Hereafter, phase-space arguments of functions will be abbreviated as
    \[
        A(\pm 1) \equiv A(\pm\vec{k}_1,\pm \vec{l}_1,\eta_1)
        \label{eq:matrix_notation1}
    \]
    and integrals over repeated arguments will be suppressed, e.g.
    \[
        [B \cdot A](1) \equiv B(1,-2) A(2) \equiv \int \dd \eta_2 \int \frac{\dd[3]{k_2}}{(2\pi)^3} \int \frac{\dd[3]{l_2}}{(2\pi)^3} \, B(1, -2) A(2).
    \label{eq:matrix_notation2}
    \]
    We define the free propagator $g_0$ in terms of its functional inverse
    \[
        g_0^{-1}(1,2) =  \qty[\partial_{\eta_1} - \vec{k}_1\cdot \nabla_{l_1} + \frac{1}{2} \, \vec{l}_1\cdot \nabla_{l_1}] \, \identity(1,2),
        \label{eq:inverse_free_propagator}
    \]
    with the identity operator
    \[
        \identity(1,2) = (2\pi)^3 \,\dirac(\vec{k}_1+\vec{k}_2) \, (2\pi)^3 \,\dirac(\vec{l}_1+\vec{l}_2) \, \dirac(\eta_1-\eta_2).
    \]
    As detailed in appendix~\ref{app:free_propagator}, the free propagator can be calculated as 
    \[ \label{eq:free_evolution}
        g_0(1,2) = (2\pi)^3 \, \dirac \qty(\vec{k}_1+\vec{k}_2) \, (2\pi)^3 \, \dirac \qty (\vec{l}_2 + \gqp(\eta_1-\eta_2) \, \vec{k}_1 + \gpp(\eta_1-\eta_2) \, \vec{l}_1)\, \Theta(\eta_1-\eta_2),
    \]
    where the functions $\gqp$ and $\gpp$ describe free particle trajectories,
    \[
        \gqp(\eta) = 2 - 2 \, \e^{-\eta/2}, \quad \gpp(\eta) = \e^{-\eta/2}.
    \]
    Furthermore, we define the vertex function
    \begin{multline}
        \gamma(1,2,3) = (2\pi)^3 \, \dirac \qty(\vec{k}_1+\vec{k}_2 +\vec{k}_3) \, \dirac(\eta_1-\eta_2)\, \dirac(\eta_1-\eta_3)\\
        \times (2\pi)^3 \, \dirac \qty(\vec{l}_1+\vec{l}_3) \, (2\pi)^3 \, \dirac \qty(\vec{l}_2)  \,\frac{3}{2 \bar{\rho}}\frac{\vec{l}_3\cdot \vec{k}_2}{k_2^2}.
    \end{multline}
    With these definitions, the Vlasov equation~\eqref{eq:Vlasov_Fourier} can be written as
    \[\label{eq:Vlasov_compact}
        g_0^{-1}(1,-2) f(2) - \gamma(1,-2,-3) f(2) f(3) = 0.
    \]
    Initial conditions on the phase-space density can be enforced by adding a term with support at initial time,
    \[
        f^\ini(1) = \dirac(\eta_1-\eta^\ini)\,f^\ini(\vec{k}_1,\vec{l}_1),
    \]
    to the right-hand side of~\eqref{eq:Vlasov_compact}.
    We split $f^\ini$ into a homogeneous background\footnote{In general, this is a volume average. Equivalently, for stochastic initial conditions, one may take the ensemble average.}~$\ev{f^\ini}$ and fluctuations~$\delta f^\ini = f^\ini-\ev{f^\ini}$, which we assume to be small.
    
    \subsection{Background--perturbation split}
    In order to establish a perturbative scheme, the phase-space distribution is split into a homogeneous background part and a perturbation
    \[
        f(1) =  \bar{f}(1) + \delta f(1).
        \label{eq:background_split}
    \]
    In principle, this split can be chosen arbitrarily. However, the choice has significant influence on the perturbative expansion.
    A trivial background is the constant comoving mean density~\cite{Valageas2001}, $\bar{f}(1) \equiv (2\pi)^3\dirac(\vec{k}_1)\,\bar{\rho}$. In this case, the dynamics are completely absorbed into the fluctuations.
    Alternatively, one could choose the average of the phase-space density, $\bar{f}(1) \equiv  \ev{f(1)}$, leading to a coupled set of equations for the (dynamical) background and the perturbations, which can be solved iteratively~\cite{Nascimento2025}.
    In this work, we take a hybrid approach, allowing for a time-dependent $\bar{f}$ while nevertheless decoupling its evolution from that of the perturbations. We choose a freely evolved background,
    \[\label{eq:free_background}
        \bar{f}(1) \equiv g_0(1,-2) \, \ev{f^\ini(2)}.
    \]
    This way, back-reactions to $\ev{f}$ are included in the dynamics of~$\delta f$. Inserting this split into the Vlasov equation,
    the dynamical equation for the fluctuations reads
    \[
        \qty[g_0^{-1}(1,-2) - \gamma(1,-2,-3) \, \bar{f}(3)] \, \delta f(2) - \gamma(1,-2,-3) \, \delta f(2) \, \delta f(3) = \delta f^\ini(1).
        \label{eq:fluctuation_eq}
    \]
    Note that for any other choice but~\eqref{eq:free_background}, an additional source term would appear. In our case, it cancels with the initial condition for the background, simplifying the perturbative expansion.
    
    \subsection{Linear perturbation theory}\label{sec:linear_theory}
    The linear part of the perturbation equation can be written as
    \[\label{eq:linear_vlasov}
        g_0^{-1}\cdot\qty[\identity-\Omega]\cdot\delta f^{(1)} = \delta f^\ini,
    \]
    where we have defined
    \[
        \Omega(1,2) &\equiv g_0(1,-3)\,\gamma(3,2,-4)\bar{f}(4) \\
        &= (2\pi)^3 \dirac(\vec{k}_1+\vec{k}_2) \, (2\pi)^3 \dirac(\vec{l}_2) \, \Tilde{\Omega}(\vec{k}_1,\vec{l}_1;\eta_1,\eta_2).
    \]
    Motivated by the Neumann series $(\identity-\Omega)^{-1} = \sum_{k=0}^{\infty} \Omega^k$ for a suitably well-behaved operator~$\Omega$, equation~\eqref{eq:linear_vlasov} can be solved in terms of the Green's function
    \[
        g = \qty[\identity + \omega] \cdot g_0 \qq{with} \omega \equiv \sum_{k=1}^{\infty} \Omega^k.
        \label{eq:linear_green_function}
    \]
    The linear correction $\omega$ to the free propagator inherits the structure of $\Omega$,
    \[
        \omega(1,2) = (2\pi)^3 \, \dirac \qty(\vec{k}_1+\vec{k}_2) \, (2\pi)^3 \dirac(\vec{l}_2)\, \Tilde{\omega}(\vec{k}_1,\vec{l}_1;\eta_1,\eta_2),
    \]
    which, by definition, can be calculated from the Volterra-type integral equation
    \[\label{eq:volterra}
        \tilde{\omega}(\vec{k},\vec{l};\eta_1,\eta_2) = \tilde{\Omega} (\vec{k},\vec{l};\eta_1,\eta_2) + \int_{\eta_2}^{\eta_1} \dd\eta \, \tilde{\Omega} (\vec{k},\vec{l};\eta_1,\eta)\, \tilde{\omega}(\vec{k},0;\eta,\eta_2).
    \]
    In practice, the self-consistency equation for $\vec{l}=0$ is solved as a first step and the general solution is obtained by an additional iteration of~\eqref{eq:volterra}.
    If $\tilde{\Omega}$ only depends on the time difference, the Volterra equation can be solved analytically using a Laplace transform.
    In general, however, the solution has to be computed numerically by discretising the time integral and solving the resulting matrix equation.
    This Volterra equation has previously been derived in the context of (resummed) kinetic field theory~\cite{Lilow2019}.
    From the linear propagator, we obtain the first-order solution
    \[
        \delta f^{(1)}(1) = g(1,-2)\,\delta f^\ini(2).
    \]
    Given that the initial fluctuations of the phase-space density were assumed to be small, linear theory motivates\footnote{We assume that momentum moments of $\delta f^\ini$ are scalars under spatial rotations and neglect vector and tensor modes. The proportionality then follows from the linearised cumulant hierarchy.} that~$\delta f^\ini \propto \delta^\ini$.
    The background split~\eqref{eq:free_background} then allows for a straightforward recursive calculation of higher-order corrections to the phase-space density, as we present in the following.
    
    \subsection{Higher-order perturbation theory}
    
    With the linear solution at hand, equation \eqref{eq:fluctuation_eq} is formally solved by
    \[
        \delta f(1) =  \delta f^{(1)}(1) + g(1,-2) \, \gamma(2,-3,-4) \, \delta f(3) \, \delta f(4).
        \label{eq:formal_solution}
    \] 
    This solution can now be iterated to obtain a perturbative expansion in powers of the initial fluctuations. In analogy to SPT, the $n$-th order correction can be expressed in terms of a symmetric integration kernel

    \[
        \delta f^{(n)}(1) \equiv \int \prod_{i=1}^n\qty[ \frac{\dd[3]{q_i}}{(2\pi)^3} \,\bar{\rho}\, \delta^\ini(\vec{q}_i) ] \, \mathcal{F}_n(1;\vec{q}_1,\dots,\vec{q}_n),
    \]
    where translation invariance ensures
    \[
        \mathcal{F}_n(1;\vec{q}_1,\dots,\vec{q}_n) \propto (2\pi)^3\,\dirac \qty(\vec{k}_1 + \vec{q}_{1\dots n}).
    \]
    Given the linear solution as computed in the previous section and, accordingly, the kernel~$\mathcal{F}_1$, higher order integration kernels can be calculated recursively from the formal solution \eqref{eq:formal_solution},
    \begin{multline}
        \mathcal{F}_n(1;\vec{q}_1,\dots, \vec{q}_n) = \\ \frac{1}{n!} \sum_{\pi \in S_n} \qty[\sum_{m=1}^{n-1} g(1,-2) \,\gamma(2,-3,-4)\, \mathcal{F}_m(3;\vec{q}_{\pi_1},\dots, \vec{q}_{\pi_m}) \, \mathcal{F}_{n-m}(4;\vec{q}_{\pi_{m+1}},\dots, \vec{q}_{\pi_n})],
        \label{eq:recursion_vlasov}
    \end{multline}
    where we explicitly symmetrised by summing over all permutations of the $\vec{q}_i$. $S_n$ denotes the symmetric group of order $n$.
    From the cumulant-generating function
    \[
        \log f(\vec{x},\vec{l},\eta) = \log\bar{f}(\vec{l},\eta) + \frac{\delta f^{(1)}(\vec{x},\vec{l},\eta)}{\bar{f}(\vec{l},\eta)} + \frac{\delta f^{(2)}(\vec{x},\vec{l},\eta)}{\bar{f}(\vec{l},\eta)} - \frac{1}{2}\qty(\frac{\delta f^{(1)}(\vec{x},\vec{l},\eta)}{\bar{f}(\vec{l},\eta)})^2 + \dots,
    \]
    perturbative corrections to momentum cumulants $\C_m$ can be calculated by taking derivatives,
    \[\label{eq:cumulant_generation}
        \C_m \equiv \im^m \underbrace{\nabla_{l}\otimes\dots\otimes\nabla_{l}}_\text{$m$ times} \log f \Big\rvert_{\vec{l}=0} = \C_m^{(0)} + \C_m^{(1)} + \C_m^{(2)} + \dots,
    \]
    e.\,g.~the density $\log\rho = \C_0$, velocity $\vec{u} = \C_1$, velocity dispersion $\sigma = \C_2$ etc. Note that the cumulants are now defined in terms of the non-canonical momentum coordinate $\vec{p}=m\dv{\vec{x}}{\eta}$.

    \subsection{Graphical representation, power spectra, and loop expansion}

    The symmetric kernels introduced in the previous section can be represented graphically in terms of tree diagrams, constructed from the building blocks
    \[
        \delta f^\ini \equiv 
        \begin{tikzpicture}[baseline=-1mm, every node/.style={scale=0.8}]
            \def\r{0.07} % radius
            \draw[thick, -{Latex[length=1mm,width=3mm/2]}] (0,0) -- (-0.5,0);
            \draw[thick,fill=white, draw=black] (0.,0.) circle (\r);
        \end{tikzpicture}, \quad
        g \equiv 
        \begin{tikzpicture}[baseline=-1mm, every node/.style={scale=0.8}]
            \draw[thick] (-0.5,0) -- (0.5,0);
            \draw[thick,-{Latex[length=1mm,width=3mm/2]}] (0.5,0) -- (-0.1,0);
        \end{tikzpicture}, \quad
        \gamma \equiv
        \begin{tikzpicture}[baseline=-1mm, every node/.style={scale=0.8}]
            \def\r{0.07} % radius
            \draw[thick, -{Latex[length=1mm,width=3mm/2]}] (0,0) -- (-0.5,0);
            \draw[thick,arrows={-Latex[length=1mm,width=3mm/2,reversed]}] (0,0) -- (30:0.5);
            \draw[thick,arrows={-Latex[length=1mm,width=3mm/2,reversed]}] (0,0) -- (-30:0.5);
            \filldraw[thick] (0.,0.) circle (\r);
        \end{tikzpicture}.
        \label{eq:tree_basic_elements}
    \]
    With these, the recursion relation \eqref{eq:recursion_vlasov} can be depicted as
    \[  \delta f^{(n)} \equiv
        \begin{tikzpicture} [baseline=0mm, every node/.style={scale=0.8}]
        \def\r{0.07} % radius
        \draw[thick] (0,0) -- (160:0.7) node[above left] {$1$};
        \draw[thick] (0,0) -- (130:0.7) node[above left] {$2$};
        \draw[thick] (0,0) -- (20:0.7) node[above right] {$n$};
        \draw[thick] (0,0) -- (50:0.7) node[above right] {$n-1$};
        \draw[thick] (0,0) -- (-90:0.4) node[above] {};
        \draw[thick, dotted] ([shift={(60:0.5)}]0,0) arc (60:120:.5);
        \draw[thick, -{Latex[length=1mm,width=3mm/2]}] (0,0) -- (0,-0.3);
        \draw[thick, -{Latex[length=1mm,width=3mm/2]}] (160:0.7) -- (160:0.3);
        \draw[thick, -{Latex[length=1mm,width=3mm/2]}] (130:0.7) -- (130:0.3);
        \draw[thick, -{Latex[length=1mm,width=3mm/2]}] (50:0.7) -- (50:0.3);
        \draw[thick, -{Latex[length=1mm,width=3mm/2]}] (20:0.7) -- (20:0.3);
        \draw[thick,fill=white, draw=black] (160:0.7) circle (\r);
        \draw[thick,fill=white, draw=black] (130:0.7) circle (\r);
        \draw[thick,fill=white, draw=black] (20:0.7) circle (\r);
        \draw[thick,fill=white, draw=black] (50:0.7) circle (\r);
        \filldraw[thick, fill=sym_vertex] (0:0) circle (.1);
    \end{tikzpicture} = \frac{1}{n!} \sum_{\pi \in S_n} \qty[\sum_{m=1}^{n-1} 
    \begin{tikzpicture}[baseline=3mm, every node/.style={scale=0.8}]
        \def\r{0.07} % radius
        \draw[thick,shift={(140:0.8)}]  (0:0.) -- (70:0.6) node[above] {$\,\pi_m$};
        \draw[thick,shift={(140:0.8)}]  (0:0.) -- (150:0.6) node[above] {$\pi_1\,$};
        \draw[thick,shift={(40:0.8)}]  (0:0.) -- (30:0.6) node[above] {$\,\pi_n$};
        \draw[thick,shift={(40:0.8)}]  (0:0.) -- (110:0.6) node[above] {$\pi_{m+1}\,$};
        \draw[thick] (0,0) -- (40:0.8);
        \draw[thick] (0,0) -- (140:0.8);
        \draw[thick] (0,0) -- (-90:0.4);
        \draw[thick, dotted] ([shift={($(140:0.8)+(80:0.4)$)}]0,0) arc (80:140:.4);
        \draw[thick, dotted] ([shift={($(40:0.8)+(40:0.4)$)}]0,0) arc (40:100:.4);
        \filldraw (0,0) circle (\r);
        \draw[thick, -{Latex[length=1mm,width=3mm/2]}] (0,0) -- (0,-0.3);
        \draw[thick, -{Latex[length=1mm,width=3mm/2]}] (40:0.8) -- (40:0.3);
        \draw[thick, -{Latex[length=1mm,width=3mm/2]}] (140:0.8) -- (140:0.3);
        \draw[thick, -{Latex[length=1mm,width=3mm/2]},shift={(140:0.8)}] (70:0.6) -- (70:0.25);
        \draw[thick, -{Latex[length=1mm,width=3mm/2]},shift={(140:0.8)}] (150:0.6) -- (150:0.25);
        \draw[thick, -{Latex[length=1mm,width=3mm/2]},shift={(40:0.8)}] (30:0.6) -- (30:0.25);
        \draw[thick, -{Latex[length=1mm,width=3mm/2]},shift={(40:0.8)}] (110:0.6) -- (110:0.25);
        \draw[thick,fill=white, draw=black,,shift={(140:0.8)}] (70:0.6) circle (\r);
        \draw[thick,fill=white, draw=black,,shift={(140:0.8)}] (150:0.6) circle (\r);
        \draw[thick,fill=white, draw=black,,shift={(40:0.8)}] (30:0.6) circle (\r);
        \draw[thick,fill=white, draw=black,,shift={(40:0.8)}] (110:0.6) circle (\r);
        \filldraw[thick, fill=sym_vertex] (140:0.8) circle (.1);
        \filldraw[thick, fill=sym_vertex] (40:0.8) circle (.1);
    \end{tikzpicture}].
    \]
    Assuming the initial density contrast~$\delta^\ini$ to be a Gaussian random field, summary statistics are calculated by taking the ensemble average. By Wick's theorem, all possible contractions of pairs of $\delta^\ini$ have to be summed over.
    Graphically, the contractions are represented by
    \[
    \langle 
        \begin{tikzpicture}
            \def\r{0.07} % radius
            \draw[thick, -{Latex[length=1mm,width=3mm/2]}] (0.1,0) -- (0.5,0);
            \draw[thick, -{Latex[length=1mm,width=3mm/2]}] (-0.1,0) -- (-0.5,0);
            \draw[thick,fill=white, draw=black] (0.1,0.) circle (\r);
            \draw[thick,fill=white, draw=black] (-0.1,0.) circle (\r);
        \end{tikzpicture}
     \rangle \equiv 
    \begin{tikzpicture}
        \def\r{0.09} % radius
        \draw[thick, -{Latex[length=1mm,width=3mm/2]}] (0,0) -- (0.5,0);
        \draw[thick, -{Latex[length=1mm,width=3mm/2]}] (0,0) -- (-0.5,0);
        \draw[thick,fill=white, draw=black] (0.,0.) circle (\r);
        \draw[thick, shift = {(0.,0.)}] ({-\r/sqrt(2)},{-\r/sqrt(2)}) -- ({\r/sqrt(2)},{\r/sqrt(2)});
        \draw[thick, shift = {(0.,0.)}] ({-\r/sqrt(2)},{\r/sqrt(2)}) -- ({\r/sqrt(2)},{-\r/sqrt(2)});
    \end{tikzpicture},
    \]
    where the crossed circle denotes an insertion of the initial power spectrum. Contributions to the one-loop power spectrum, for example, are given in terms of the connected diagrams
    \[ \label{eq:one-loop-spectra}
    \ev{\delta f^{(3)} \delta f^{(1)}}_\text{connected} &= 3\,
        \begin{tikzpicture}[baseline=-1mm, every node/.style={scale=0.8}]
            \def\r{0.09} % radius
            \draw[thick] (0,0) -- (1.2,0);
            \draw[thick] (0,0) -- (-0.6,0);
            \draw[thick] (0,0) arc (-90:90:0.4);
            \draw[thick, -{Latex[length=1mm,width=3mm/2]},shift = {(0.4,0.4)}] (0,0) -- (0,-0.05);
            \begin{scope}[xscale=-1]
                \draw[thick] (0,0) arc (-90:90:0.4);
                \draw[thick, -{Latex[length=1mm,width=3mm/2]},shift = {(0.4,0.4)}] (0,0) -- (0,-0.05);
             \end{scope}
            \draw[thick, -{Latex[length=1mm,width=3mm/2]}] (0.7,0) -- (1.1,0);
            \draw[thick, -{Latex[length=1mm,width=3mm/2]}] (0.7,0) -- (0.3,0);
            \draw[thick, -{Latex[length=1mm,width=3mm/2]}] (0.,0) -- (-0.5,0);
            \draw[thick,fill=white, draw=black] (0.7,0) circle (\r);
            \draw[thick, shift = {(0.7,0)}] ({-\r/sqrt(2)},{-\r/sqrt(2)}) -- ({\r/sqrt(2)},{\r/sqrt(2)});
            \draw[thick, shift = {(0.7,0)}] ({-\r/sqrt(2)},{\r/sqrt(2)}) -- ({\r/sqrt(2)},{-\r/sqrt(2)});
            \draw[thick,fill=white, draw=black] (0.,.8) circle (\r);
            \draw[thick, shift = {(0.,.8)}] ({-\r/sqrt(2)},{-\r/sqrt(2)}) -- ({\r/sqrt(2)},{\r/sqrt(2)});
            \draw[thick, shift = {(0.,.8)}] ({-\r/sqrt(2)},{\r/sqrt(2)}) -- ({\r/sqrt(2)},{-\r/sqrt(2)});
            \filldraw[thick, fill=sym_vertex] (0:0) circle (.1);
        \end{tikzpicture} = 3\,
        \begin{tikzpicture}[baseline=-1mm, every node/.style={scale=0.8}]
            \def\r{0.09} % radius
            \filldraw (0:0) circle (0.07);
            \filldraw (0,0.4) circle (0.07);
            \draw[thick] (0,0) -- (1.2,0);
            \draw[thick] (0,0) -- (-0.6,0);
            \draw[thick] (0,0) --(0,0.4);
            \draw[thick] (0,0.4) arc (-90:90:0.2);
            \draw[thick, -{Latex[length=1mm,width=3mm/2]},shift = {(0.2,0.6)}] (0,0) -- (0,-0.05);
            \begin{scope}[xscale=-1]
                \draw[thick] (0,0.4) arc (-90:90:0.2);
                \draw[thick, -{Latex[length=1mm,width=3mm/2]},shift = {(0.2,0.6)}] (0,0) -- (0,-0.05);
             \end{scope}
            \draw[thick, -{Latex[length=1mm,width=3mm/2]}] (0.7,0) -- (1.1,0);
            \draw[thick, -{Latex[length=1mm,width=3mm/2]}] (0.7,0) -- (0.3,0);
            \draw[thick, -{Latex[length=1mm,width=3mm/2]}] (0.,0) -- (-0.5,0);
            \draw[thick, -{Latex[length=1mm,width=3mm/2]}] (0.,0.4) -- (0.,0.15);
            \draw[thick,fill=white, draw=black] (0.7,0) circle (\r);
            \draw[thick, shift = {(0.7,0)}] ({-\r/sqrt(2)},{-\r/sqrt(2)}) -- ({\r/sqrt(2)},{\r/sqrt(2)});
            \draw[thick, shift = {(0.7,0)}] ({-\r/sqrt(2)},{\r/sqrt(2)}) -- ({\r/sqrt(2)},{-\r/sqrt(2)});
            \draw[thick,fill=white, draw=black] (0.,.8) circle (\r);
            \draw[thick, shift = {(0.,.8)}] ({-\r/sqrt(2)},{-\r/sqrt(2)}) -- ({\r/sqrt(2)},{\r/sqrt(2)});
            \draw[thick, shift = {(0.,.8)}] ({-\r/sqrt(2)},{\r/sqrt(2)}) -- ({\r/sqrt(2)},{-\r/sqrt(2)});
        \end{tikzpicture} + 6 \,
        \begin{tikzpicture}[baseline=-1mm, every node/.style={scale=0.8}]
            \def\r{0.09} % radius
            \draw[thick] (0,0) arc (0:-180:0.4);
            \draw[thick] (0,0) arc (0:180:0.4);
            \draw[thick] (0,0) -- (0.8,0.);
            \draw[thick] (-.8,0) -- (-1.2,0.);
            \filldraw (0:0) circle (0.07);
            \filldraw (180:0.8) circle (0.07);
            \draw[thick, -{Latex[length=1mm,width=3mm/2]}] (0.4,0.) -- (0.7,0.);
            \draw[thick, -{Latex[length=1mm,width=3mm/2]}] (0.4,0.) -- (0.1,0.);
            \draw[thick, -{Latex[length=1mm,width=3mm/2]}] (-0.8,0.) -- (-1.1,0.);
            \draw[thick, -{Latex[length=1mm,width=3mm/2]},shift={($(-0.4,0)+(135:0.4)$)}] (0,0.) -- (-135:0.1);
            \draw[thick, -{Latex[length=1mm,width=3mm/2]},shift={($(-0.4,0)+(45:0.4)$)}] (0,0.) -- (-45:0.1);
            \draw[thick, -{Latex[length=1mm,width=3mm/2]},shift={(-0.4,-0.4)}] (0,0.) -- (-.04,0.);
            \draw[thick,fill=white, draw=black] (-0.4,0.4) circle (\r);
            \draw[thick, shift = {(-0.4,0.4)}] ({-\r/sqrt(2)},{-\r/sqrt(2)}) -- ({\r/sqrt(2)},{\r/sqrt(2)});
            \draw[thick, shift = {(-0.4,0.4)}] ({-\r/sqrt(2)},{\r/sqrt(2)}) -- ({\r/sqrt(2)},{-\r/sqrt(2)});
            \draw[thick,fill=white, draw=black] (0.4,0.) circle (\r);
            \draw[thick, shift = {(0.4,0.)}] ({-\r/sqrt(2)},{-\r/sqrt(2)}) -- ({\r/sqrt(2)},{\r/sqrt(2)});
            \draw[thick, shift = {(0.4,0.)}] ({-\r/sqrt(2)},{\r/sqrt(2)}) -- ({\r/sqrt(2)},{-\r/sqrt(2)});
        \end{tikzpicture}, \\
        \ev{\delta f^{(2)} \delta f^{(2)}}_\text{connected} &= 2\,
        \begin{tikzpicture}[baseline=-1mm, every node/.style={scale=0.8}]
            \def\r{0.09} % radius
            \draw[thick] (0,0) arc (0:-180:0.4);
            \draw[thick] (0,0) arc (0:180:0.4);
            \draw[thick] (0,0) -- (0.4,0.);
            \draw[thick] (-.8,0) -- (-1.2,0.);
            \draw[thick, -{Latex[length=1mm,width=3mm/2]}] (-0.8,0.) -- (-1.1,0.);
            \draw[thick, -{Latex[length=1mm,width=3mm/2]}] (0.,0.) -- (0.3,0.);
            \draw[thick, -{Latex[length=1mm,width=3mm/2]},shift={($(-0.4,0)+(135:0.4)$)}] (0,0.) -- (-135:0.1);
            \draw[thick, -{Latex[length=1mm,width=3mm/2]},shift={($(-0.4,0)+(45:0.4)$)}] (0,0.) -- (-45:0.1);
            \draw[thick, -{Latex[length=1mm,width=3mm/2]},shift={($(-0.4,0)+(-135:0.4)$)}] (0,0.) -- (135:0.1);
            \draw[thick, -{Latex[length=1mm,width=3mm/2]},shift={($(-0.4,0)+(-45:0.4)$)}] (0,0.) -- (45:0.1);
            \draw[thick,fill=white, draw=black] (-0.4,0.4) circle (\r);
            \draw[thick, shift = {(-0.4,0.4)}] ({-\r/sqrt(2)},{-\r/sqrt(2)}) -- ({\r/sqrt(2)},{\r/sqrt(2)});
            \draw[thick, shift = {(-0.4,0.4)}] ({-\r/sqrt(2)},{\r/sqrt(2)}) -- ({\r/sqrt(2)},{-\r/sqrt(2)});
            \draw[thick,fill=white, draw=black] (-0.4,-0.4) circle (\r);
            \draw[thick, shift = {(-0.4,-0.4)}] ({-\r/sqrt(2)},{-\r/sqrt(2)}) -- ({\r/sqrt(2)},{\r/sqrt(2)});
            \draw[thick, shift = {(-0.4,-0.4)}] ({-\r/sqrt(2)},{\r/sqrt(2)}) -- ({\r/sqrt(2)},{-\r/sqrt(2)});
            \filldraw[thick, fill=sym_vertex] (0:0) circle (.1);
            \filldraw[thick, fill=sym_vertex] (180:0.8) circle (.1);
        \end{tikzpicture} = 2 \,
        \begin{tikzpicture}[baseline=-1mm, every node/.style={scale=0.8}]
            \def\r{0.09} % radius
            \draw[thick] (0,0) arc (0:-180:0.4);
            \draw[thick] (0,0) arc (0:180:0.4);
            \draw[thick] (0,0) -- (0.4,0.);
            \draw[thick] (-.8,0) -- (-1.2,0.);
            \filldraw (0:0) circle (0.07);
            \filldraw (180:0.8) circle (0.07);
            \draw[thick, -{Latex[length=1mm,width=3mm/2]}] (-0.8,0.) -- (-1.1,0.);
            \draw[thick, -{Latex[length=1mm,width=3mm/2]}] (0.,0.) -- (0.3,0.);
            \draw[thick, -{Latex[length=1mm,width=3mm/2]},shift={($(-0.4,0)+(135:0.4)$)}] (0,0.) -- (-135:0.1);
            \draw[thick, -{Latex[length=1mm,width=3mm/2]},shift={($(-0.4,0)+(45:0.4)$)}] (0,0.) -- (-45:0.1);
            \draw[thick, -{Latex[length=1mm,width=3mm/2]},shift={($(-0.4,0)+(-135:0.4)$)}] (0,0.) -- (135:0.1);
            \draw[thick, -{Latex[length=1mm,width=3mm/2]},shift={($(-0.4,0)+(-45:0.4)$)}] (0,0.) -- (45:0.1);
            \draw[thick,fill=white, draw=black] (-0.4,0.4) circle (\r);
            \draw[thick, shift = {(-0.4,0.4)}] ({-\r/sqrt(2)},{-\r/sqrt(2)}) -- ({\r/sqrt(2)},{\r/sqrt(2)});
            \draw[thick, shift = {(-0.4,0.4)}] ({-\r/sqrt(2)},{\r/sqrt(2)}) -- ({\r/sqrt(2)},{-\r/sqrt(2)});
            \draw[thick,fill=white, draw=black] (-0.4,-0.4) circle (\r);
            \draw[thick, shift = {(-0.4,-0.4)}] ({-\r/sqrt(2)},{-\r/sqrt(2)}) -- ({\r/sqrt(2)},{\r/sqrt(2)});
            \draw[thick, shift = {(-0.4,-0.4)}] ({-\r/sqrt(2)},{\r/sqrt(2)}) -- ({\r/sqrt(2)},{-\r/sqrt(2)});
        \end{tikzpicture}.
    \]
    In the last step, the recursion relations have been traced back to $\mathcal{F}_1$, such that the diagrams can be decomposed into the fundamental building blocks \eqref{eq:tree_basic_elements}.
    Due to the perturbative ansatz for the background phase-space density, the loop expansion includes tadpole diagrams encoding corrections to the mean phase-space density $\ev{f}$.
    
    The perturbative scheme in terms of the recursion relation~\eqref{eq:recursion_vlasov} is efficient and reminiscent of the recursive approach to SPT. In particular, we do not have to deal explicitly with back-reaction effects, as they are contained in the kernels, contrary to ref.~\cite{Nascimento2025}. Alternatively, ref.~\cite{Daus2025PhD} develops a path-integral formulation of the Klimontovich equation, producing a similar diagrammatic expansion to the right-hand side of~\eqref{eq:one-loop-spectra}. Again, the integration kernels presented here summarize the effects of multiple loop diagrams. While the number of kernels grows linearly with the perturbative order, the number of ``decomposed'' loop diagrams grows factorially.
        
    \section{Perfectly cold initial conditions}\label{sec:perfectly_cold}

    The formalism presented in the previous section works independently of the explicit structure of initial conditions. In this section, perfectly cold initial conditions (CICs) are considered. Ignoring vorticity and assuming that linear SPT is a valid description at the initial time, the initial phase-space density reads
    \[ \label{eq:cold_ics}
        f^{\ini}(\vec{k},\vec{l}) = (2\pi)^3 \dirac(\vec{k})\,\bar{\rho} + \bar{\rho}\, \qty(1 + \frac{\vec{l}\cdot \vec{k}}{k^2}) \delta^\ini(\vec{k}),
    \]
    where the growing mode has been selected. Note that for these initial conditions, the background split~\eqref{eq:free_background} exactly coincides with that of ref.~\cite{Valageas2001}.
    The Volterra equation~\eqref{eq:volterra} can be solved analytically using a Laplace transform in $\eta_1-\eta_2$,
    \[\label{eq:omega_cold}
        \tilde\omega(\vec{k},\vec{l};\eta_1,\eta_2) = \frac{3}{5} \qty[ \e^{\eta_1-\eta_2} \qty(1+ \frac{\vec{k}\cdot \vec{l}}{k^2}) - \e^{-\frac{3}{2}(\eta_1-\eta_2)} \qty(1 - \frac{3}{2} \frac{\vec{k}\cdot \vec{l}}{k^2}) ] \Theta(\eta_1-\eta_2).
    \]
    The linear solution is given by
    \[
        \delta f^{(1)}(1) = g(1,-2) \,\delta f^\ini(2) = \bar{\rho} \, \e^{\eta_1} \qty(1+ \frac{\vec{k}_1\cdot \vec{l}_1}{k_1^2}) \delta^\ini(\vec{k}_1),
    \label{eq:linear_sol}
    \]
    recovering, in particular, the linear growth of density perturbations, $\delta^{(1)} = \e^{\eta_1}\,\delta^\ini$. In general, the nonlinear corrections contain terms growing or decaying at different rates. At late times, all but the fastest-growing contribution can safely be neglected.
    Up to $\mathcal{F}_3$, we have checked explicitly\footnote{While for low orders it might be feasible to do the calculation by hand, we use SageMath~\cite{Sage} to perform symbolic calculations for higher-order kernels.} that the late-time results of SPT for all momentum moments are recovered. In one spatial dimension, we find a closed form solution for the growing-mode Vlasov kernels of arbitrary order,
    \[\label{eq:1d_solution}
        \mathcal{F}_n(1; q_1,\dots,q_n) = 2\pi \dirac\qty(k_1+q_{1\dots n})\,\bar{\rho}^{1-n}\,\e^{n\eta_1}\,\frac{(q_{1\dots n} - l_1)^n}{n! \prod_{j=1}^n q_j},
    \]
    which we prove by induction in appendix~\ref{app:proof_1D}, using the recursion relation~\eqref{eq:recursion_vlasov}. For $l_1=0$, this reduces to the known solutions for the density kernels of SPT/LPT~\cite{Grinstein1987,McQuinn2016}, and similarly for the velocity kernels by first taking appropriate derivatives with respect to $l_1$, in accordance with eq.~\eqref{eq:cumulant_generation}.
    In fact, also for the three-dimensional system, the perturbative equivalence to SPT should hold to all orders. This has been argued for the first time in ref.~\cite{Valageas2001}. While the full Vlasov equation allows for dynamical deviation from CICs, the SSA poses a fixed point of the momentum-cumulant hierarchy, since the third and higher momentum cumulants are sourceless in this case~\cite{Pueblas2009}. This apparent contradiction between the Vlasov equation and its cumulant hierarchy is rooted in the divergence of the density contrast at stream crossing in the case of CICs. Mathematically, the hierarchy is no longer valid thereafter.
    Perturbation theory corresponds to an iterative solution of the momentum-cumulant hierarchy and suffers from the aforementioned fixed point. However, deviating slight from CICs avoids the problematic singularities. As a consequence, arbitrary momentum cumulants are generated at any order, which will be made explicit in the subsequent section.
    
    \section{Beyond the single-stream approximation}\label{sec:slightly_warm}

    Recently, there have been several approaches to extend SPT beyond the SSA, relying on higher truncations of the cumulant hierarchy~\cite{Buchert1997,McDonald2011,Erschfeld2019,Erschfeld2024,Garny2023lin,Garny2023nonlin,Garny2025a,Garny2025}. Including velocity dispersion has a significant qualitative effect, namely a suppression of structure on small scales. Truncations beyond the second cumulant seem to influence the results mostly quantitatively~\cite{Garny2023lin}.
    The approach presented in this work, on the other hand, does not assume any truncation of the Vlasov momentum-cumulant hierarchy.
    As the SSA poses the only non-trivial fixed point of the latter, deviating only slightly from CICs breaks the aforementioned self-consistency and generates non-vanishing contributions to \emph{all} momentum cumulants. In order to demonstrate this qualitatively, we consider a simple extension of the initial conditions~\eqref{eq:cold_ics} by including 
    a homogeneous and isotropic initial velocity dispersion $\sigma^\ini$,
    \[\label{eq:warm_ICs}
        f^\ini(\vec{k},\vec{l}) = \bar{\rho} \qty[(2\pi)^3\dirac(\vec{k})+ \qty(1 + \frac{\vec{k}\cdot \vec{l}}{k^2}) \delta^\ini(\vec{k})] \exp(-\frac{\sigma^\ini l^2}{2}).
    \]
    In a more realistic model, one could include fluctuations of the velocity dispersion and also non-vanishing higher momentum cumulants to these initial conditions. The perturbative calculations of this work can be performed in exactly the same manner in this case. However, we choose to consider only the simple form~\eqref{eq:warm_ICs} in the following, given that it already yields many important nontrivial effects while introducing only a single additional parameter.
    
    \subsection{Linear perturbation theory}
\begin{figure}
        \centering
        \includegraphics[width=.75\linewidth]{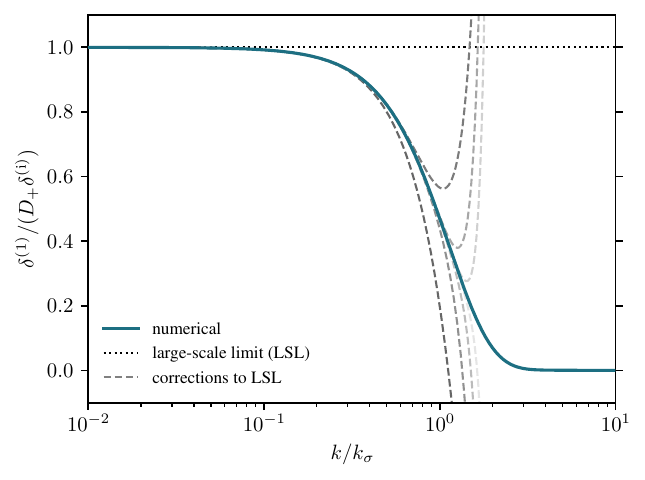}
        \caption{\label{fig:density} Late-time linear density contrast with non-vanishing initial velocity dispersion, normalized to the perfectly cold case. The solid blue line corresponds to the numerical solution. Dashed grey lines indicate corrections to the LSL (black, dotted) up to seventh order in $(k/k_\sigma)^2$.}
    \end{figure}

\begin{figure}
        \centering
        \includegraphics[width=.75\linewidth]{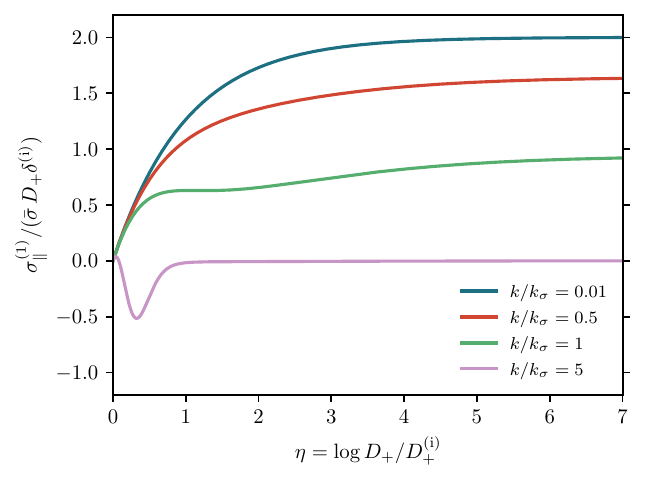}
        \caption{\label{fig:sigma} Numerical results for the time evolution of the longitudinal mode of the linear velocity dispersion relative to the background, for four different scales.}
    \end{figure}

    In contrast to the perfectly cold case, the Volterra equation~\eqref{eq:volterra} can no longer be solved analytically, because $\tilde\Omega$ does not depend solely on the time difference, $\eta_{12} = \eta_1-\eta_2$,
    \[
        \tilde\Omega(\vec{k},\vec{l};\eta_1,\eta_2) &= \frac{3}{2} \qty[\gqp(\eta_{12}) + \gpp(\eta_{12}) \frac{\vec{k}\cdot\vec{l}}{k^2}] \exp(-\frac{\bar\sigma(\eta_2)}{2} \big\lvert\gqp(\eta_{12})\,\vec{k} + \gpp(\eta_{12})\,\vec{l}\big\rvert^2),
    \]
    where $\bar{\sigma}$ is the (freely evolved) background velocity dispersion
    \[
        \bar{\sigma}(\eta)\,\identity_3 \equiv -(\nabla_{l}\otimes\nabla_{l}) \log\bar{f}(\vec{l},\eta)\Big\rvert_{\vec{l}=0} = \gpp(\eta)^2\,\sigma^\ini \identity_3,
    \]
    with the three-dimensional unit matrix $\identity_3$.
    Instead, we discretise time and solve the corresponding matrix equation numerically. For popular dark matter candidates, like WIMPs, the initial velocity dispersion is small compared to the scales of interest~\cite{Boyanovsky2011}, $k/k_\sigma\ll 1$, with~$k_\sigma = 1/\sqrt{\sigma^\ini}$. In addition to the numerical solution, we may therefore perform a large-scale expansion in powers of $k/k_\sigma$. Formally, we write
    \[\label{eq:large_scale_exp}
        \Omega = \sum_{k=0}^{\infty} (\sigma^\ini)^k\,\Omega_k,\quad \omega = \sum_{k=0}^{\infty} (\sigma^\ini)^k\,\omega_k.
    \]
    Note that $\omega_0$ is precisely the perfectly cold solution~\eqref{eq:omega_cold}, which is recovered in the large-scale limit (LSL). Higher-order corrections can be calculated recursively,
    \[
        \omega_k = (\identity+\omega_0) \cdot \qty[\Omega_k + \sum_{j=0}^{k-1} \Omega_{k-j} \cdot \omega_{j}].
    \]
    In figure~\ref{fig:density}, we show the late-time linear density contrast $\delta^{(1)}(\vec{k},\eta) = \delta f^{(1)}(\vec{k},0,\eta)/\bar{\rho}$.
    On sufficiently small scales, $k \gtrsim k_\sigma$, it is exponentially suppressed compared to linear SPT---small-scale structure diffuses due to the non-vanishing velocity dispersion. This is a consequence of our background--perturbation split, which includes the dispersion due to free streaming nonperturbatively. At higher orders, this small-scale damping might regulate the UV divergences that SPT suffers from.
    We complement our numerical results with approximations up to seventh order in the aforementioned expansion\footnote{When including a longitudinal scalar perturbation of the velocity dispersion in the initial conditions as done in ref.~\cite{McDonald2011}, we exactly reproduce their coefficients up to order $(k/k_\sigma)^2$. Note, however, that ref.~\cite{McDonald2011} uses a truncation of the momentum-cumulant hierarchy.},
    \begin{multline}\label{eq:ltls_expansion}
       \frac{\delta^{(1)}(\vec{k},\eta\gg 0)}{D_+(\eta)\,\delta^\ini(\vec{k})} \approx 1 - \frac{4}{5} \qty(\frac{k}{k_\sigma})^2 + \frac{64}{175} \qty(\frac{k}{k_\sigma})^4 - \frac{466}{3675} \qty(\frac{k}{k_\sigma})^6 + \frac{3203}{86625} \qty(\frac{k}{k_\sigma})^8 \\
       - \frac{195724}{20645625} \qty(\frac{k}{k_\sigma})^{10}
       + \frac{1290913336}{591806840625} \qty(\frac{k}{k_\sigma})^{12} - \frac{14574682789}{32011370015625} \qty(\frac{k}{k_\sigma})^{14},
    \end{multline}
    matching the numerical result closely for $k/k_\sigma \lesssim 1$.
    Due to our choice of initial conditions, the velocity dispersion can be decomposed into two scalar modes within linear theory,
    \[
       \sigma^{(1)}(\vec{k},\eta) = \left.-(\nabla_{l}\otimes\nabla_{l})\frac{\delta f^{(1)}(\vec{k},\vec{l},\eta)}{\bar{f}(\vec{l},\eta)}\right\rvert_{\vec{l}=0} \equiv \sigma^{(1)}_\identity(\vec{k},\eta)\, \identity_3 + \sigma^{(1)}_\parallel(\vec{k},\eta)\,\frac{\vec{k}\otimes\vec{k}}{k^2}.
    \]
    In the following discussion, we will consider dimensionless quantities, measuring the growth of perturbations relative to the background. This also allows for a direct comparison to results obtained from the truncated Vlasov hierarchy \cite{McDonald2011, Garny2023lin, Garny2023nonlin}.
    The mode $\sigma^{(1)}_\identity/\bar{\sigma}$ decays, whereas the longitudinal mode grows linearly at late times, $\sigma^{(1)}_\parallel/\bar{\sigma} \propto D_+$, as presented in figure~\ref{fig:sigma}. This is in qualitative agreement with the findings of refs.~\cite{McDonald2011, Garny2023lin}.
    We find that the density contrast, the divergence of the velocity and the background-normalised longitudinal velocity dispersion all obey the same large-scale expansion~\eqref{eq:ltls_expansion},
    \[
       \delta^{(1)} = \theta^{(1)} = \frac{1}{2}\frac{\sigma^{(1)}_\parallel}{\bar{\sigma}}, \qq{for} \eta\gg0.
    \]
    This equivalence also seems to hold for the numerical results. 
    Longitudinal contributions to all higher momentum cumulants are also generated,
    \[
        \C_m^{(1)}(\vec{k},\eta) \equiv c_m^{(1)}(\vec{k},\eta)\,\im^m\vec{k}^{\otimes m}
        \begin{cases*}
            k^{-m} & $m$ even\\
            k^{-(m+1)} & $m$ odd
        \end{cases*}.
    \]
    At late times and leading order in $k/k_\sigma$, the dimensionless quantities for the third to eighth cumulants read
    \[
        \frac{c_3^{(1)}(\vec{k},\eta)}{\bar{\sigma}(\eta)\,\delta^\ini(\vec{k})} &\approx 12\,\qty(\frac{k}{k_\sigma})^2,  &
        \frac{c_4^{(1)}(\vec{k},\eta)}{\bar{\sigma}^2(\eta)\,\delta^\ini(\vec{k})} &\approx -48\,\log(D_+(\eta)) \,\qty(\frac{k}{k_\sigma})^2, \\
        \frac{c_5^{(1)}(\vec{k},\eta)}{\bar{\sigma}^2(\eta)\,\delta^\ini(\vec{k})} &\approx -\frac{480}{\sqrt{D_+(\eta)}} \,\qty(\frac{k}{k_\sigma})^4, &
        \frac{c_6^{(1)}(\vec{k},\eta)}{\bar{\sigma}^3(\eta)\,\delta^\ini(\vec{k})} &\approx -480\,\qty(\frac{k}{k_\sigma})^4, \\
        \frac{c_7^{(1)}(\vec{k},\eta)}{\bar{\sigma}^3(\eta)\,\delta^\ini(\vec{k})} &\approx \frac{896}{\sqrt{D_+(\eta)}} \,\qty(\frac{k}{k_\sigma})^6, &
        \frac{c_8^{(1)}(\vec{k},\eta)}{\bar{\sigma}^4(\eta)\,\delta^\ini(\vec{k})} &\approx -1792\,\qty(\frac{k}{k_\sigma})^6.
    \]
    Higher cumulants are suppressed by larger powers of $k/k_\sigma$. Excluding the fourth cumulant, none of the above have a growing contribution. We expect this to extend to arbitrary higher cumulants at linear order in perturbation theory. Thus, their physical significance is minute and we only consider them for qualitative purposes, demonstrating that they are generated at all.
    Going to higher orders in $k/k_\sigma$, the third and fourth momentum cumulants are also damped at the same scale as the density contrast, i.\,e.~admit the same large-scale expansion~\eqref{eq:ltls_expansion} up to a time-dependent factor, for $\eta\gg0$. This is no longer true for the fifth and higher cumulants.
     
    \subsection{Second-order perturbation theory}
    In order to discuss some qualitative effects at second perturbative order, we restrict ourselves to the large-scale expansion~\eqref{eq:large_scale_exp}, given that it provides an accurate description for scales $k\lesssim k_\sigma$.
    A first difference to the linear approximation is that $\sigma$ acquires non-longitudinal contributions,
    \begin{multline}
        \frac{\sigma^{(2)}(\vec{k},\eta)}{\bar{\sigma}(\eta)} \approx D_+^2(\eta)\int\frac{\dd[3]q_1\,\dd[3]q_2}{(2\pi)^3}\,\dirac(\vec{k}+\vec{q}_1+\vec{q}_2)\, \delta^\ini(\vec{q}_1) \delta^\ini(\vec{q}_2) \\
        \times\qty{\qty[\frac{3}{7}+\frac{\vec{q}_1\cdot\vec{q}_2}{q_1 q_2} \qty(\frac{q_1}{q_2}+\frac{q_2}{q_1}) + \frac{11}{7}\frac{(\vec{q}_1\cdot\vec{q}_2)^2}{q_1^2 q_2^2}] \frac{\vec{k}\otimes\vec{k}}{k^2} + \frac{\vec{q}_1\cdot\vec{q}_2}{q_1 q_2}\frac{\vec{q}_1\otimes\vec{q}_2}{q_1 q_2} },
    \end{multline}
    at leading order in $k/k_\sigma$ and for late times. In particular, the last term in the above expression contains an anisotropic stress component, which, in turn, acts as a source for vorticity at higher orders. For the third cumulant, we have
    \begin{multline}
        \frac{\C_3^{(2)}(\vec{k},\eta)}{\bar{\sigma}(\eta)} \approx -\im\,D_+(\eta)\qty(\frac{k}{k_\sigma})^2 \int\frac{\dd[3]q_1\,\dd[3]q_2}{(2\pi)^3}\,\dirac(\vec{k}+\vec{q}_1+\vec{q}_2)\, \delta^\ini(\vec{q}_1) \delta^\ini(\vec{q}_2) \\
        \times \Bigg\{6\,\qty[\frac{1}{7}+\frac{\vec{q}_1\cdot\vec{q}_2}{q_1 q_2} \qty(\frac{q_1}{q_2}+\frac{q_2}{q_1}) + \frac{13}{7}\frac{(\vec{q}_1\cdot\vec{q}_2)^2}{q_1^2 q_2^2}] \frac{\vec{k}\otimes\vec{k}\otimes\vec{k}}{k^4}
        -6\,\frac{\vec{q}_1\cdot\vec{q}_2}{q_1 q_2}\frac{\vec{q}_1\otimes\vec{q}_2\otimes\vec{q}_2}{q_1 q_2 k^2}\\
         - \frac{27}{100}\qty[\frac{10}{7}+\frac{\vec{q}_1\cdot\vec{q}_2}{q_1 q_2} \qty(\frac{q_1}{q_2}+\frac{q_2}{q_1}) + \frac{4}{7}\frac{(\vec{q}_1\cdot\vec{q}_2)^2}{q_1^2 q_2^2}]\frac{\vec{q}_1\otimes\identity_3}{k^2}\Bigg\}_\text{sym},
    \end{multline}
    where we fully symmetrize all tensors. In contrast to the linear approximation, the second-order contribution grows relative to the background. This is to be expected by the construction of our perturbative expansion, where higher orders grow increasingly fast. This way, growing contributions to \emph{all} momentum cumulants arise when deviating from CICs. While the initial conditions~\eqref{eq:warm_ICs} only constitute a very simple extension, they serve as a demonstration that a perturbative treatment of the Vlasov hierarchy without any truncation is feasible.

    \section{Conclusions and outlook}\label{sec:conclusions}
    In this work, we established a perturbative approach to the full Vlasov equation that does not depend on any truncation of its cumulant hierarchy. Similarly to refs.~\cite{Valageas2001,Nascimento2025}, it is based on splitting the phase-space density into a background and fluctuations. In contrast to earlier works, however, we choose the former to be freely evolving. This way, the trivial part of the dynamics is absorbed into the background while all interaction effects remain in the perturbations. At linear order, the fluctuations are obtained by solving the Volterra-type integral equation~\eqref{eq:volterra}. Higher order corrections can be computed recursively, using eq.~\eqref{eq:recursion_vlasov}.
    For perfectly cold initial conditions, we exactly reproduce the SPT kernels up to third order. In the one-dimensional case, we show that this extends to arbitrary orders by providing the closed-form solution~\eqref{eq:1d_solution} for the phase-space kernels. This is consistent with the findings of ref.~\cite{Valageas2001}. In fact, the equivalence is to be expected, since perturbative approaches to the Vlasov equation correspond to an iterative solution of its cumulant hierarchy, for which the SSA poses a self-consistent truncation~\cite{Pueblas2009}. Therefore, starting from CICs, no higher cumulants can be generated perturbatively. Note that this is an issue of perturbation theory, or, consequently, the transition from the Vlasov equation to its momentum-cumulant hierarchy. Due to stream crossing, momentum moments like the mass density may formally diverge. While the hierarchy becomes invalid, exact solutions of the Vlasov equation capture all effects of stream-crossing even for CICs.
    In perturbation theory, the aforementioned single-stream fixed point can be avoided by deviating from CICs. Here, we considered a simple extension by including a non-vanishing homogeneous and isotropic initial velocity dispersion. In this case, the Volterra equation for the linear solution can no longer be solved analytically. In addition to the numerical solution, we consider a large-scale expansion relative to the scale set by the initial velocity dispersion. On scales smaller than the latter, structure is strongly suppressed, which might regulate the UV divergences that SPT suffers from at higher perturbative orders. As a consequence of deviating from CICs, all momentum cumulants are dynamically generated and, at sufficiently high order in perturbation theory, grow with respect to the background. In addition, the velocity dispersion develops an anisotropic stress component at second order, providing a source for vorticity at higher orders. While these are only qualitative results, they serve as a demonstration that perturbative approaches to the Vlasov equation beyond the SSA are possible without the need for a truncation of the momentum-cumulant hierarchy.
    
    Our method has similarities to other approaches beyond the SSA. Refs.~\cite{Buchert1997,McDonald2011,Erschfeld2019,Erschfeld2024,Garny2023lin,Garny2023nonlin,Garny2025a,Garny2025}, which do rely on such a truncation, also deviate from CICs by introducing a non-vanishing initial velocity dispersion. In refs.~\cite{Garny2023lin,Garny2023nonlin}, perturbation theory is performed around an a priori unspecified background. For the special case of scale symmetry, the latter is derived self-consistently. In our framework, the Volterra equation~\eqref{eq:volterra} can be solved given an arbitrary homogeneous background for the phase-space density. Considering that the choice of truncation influences results quantitatively~\cite{Garny2023lin} and that it can be avoided in our approach, finding a suitable background field could allow for a more efficient and accurate expansion.
    In ref.~\cite{McDonald2011}, a quickly growing homogeneous velocity dispersion is obtained using renormalisation group techniques. A similar observation was made in ref.~\cite{Erschfeld2024}, making use of truncated Dyson-Schwinger equations.
    It is expected that a purely perturbative treatment will not capture all relevant physical effects. In particular, it cannot fully describe the formation of (quasi-)\allowbreak bound structures, like dark matter halos, and their subsequent decoupling from large-scale dynamics.
    The present work provides the basis for applying similar nonperturbative techniques without employing a truncation of the momentum-cumulant hierarchy.
    We leave such considerations for future work.

    \acknowledgments
    We thank Marie-Ann Grebe for proofreading and commenting on this manuscript.
    This work was supported by the Deutsche Forschungsgemeinschaft (DFG, German Research Foundation) under Germany's Excellence Strategy EXC 2181/1 - 390900948 (the Heidelberg STRUCTURES Excellence Cluster). HH is supported by the Deutsche Forschungsgemeinschaft (DFG, German Research Foundation) - 528166846.
    MS is supported by Stiftung der Deutschen Wirtschaft (sdw).
    
    \appendix
    
    \section{Calculating the free propagator}\label{app:free_propagator}
    With the time coordinate defined in section~\ref{sec:perturbation_theory}, the \emph{free} Hamiltonian equations of motion in the EdS approximation read
    \[
        \dv{}{\eta}\!\begin{pmatrix}
            \vec{x} \\
            \vec{p}
        \end{pmatrix}
        = \begin{pmatrix}
            \vec{p}/m\\
            -\vec{p}/2
        \end{pmatrix}.
    \]
    They are solved in terms of a Green's function,
    \[
        \begin{pmatrix}
            \vec{x}(\eta) \\ \vec{p}(\eta)
        \end{pmatrix} = \begin{pmatrix}
            1 & \gqp(\eta) \\
            0 & \gpp(\eta)
        \end{pmatrix} \begin{pmatrix}
            \vec{x}^\ini \\ \vec{p}^\ini
        \end{pmatrix}, \quad \gqp(\eta) = \frac{1}{m}\qty(2 - 2 \, \e^{-\eta/2}), \quad \gpp(\eta) = \e^{-\eta/2}.
    \]
    Given that the Vlasov equation is a conservation equation for the phase-space density, its free evolution is therefore described by the propagator
    \[
        g_0(\vec{x}_1,\vec{p}_1,\eta_1,\vec{x}_2,\vec{p}_2,\eta_2) = \dirac\qty[
        \begin{pmatrix}
            \vec{x}_1 \\ \vec{p}_1
        \end{pmatrix}
        - \begin{pmatrix}
            1 & \gqp(\eta_1-\eta_2) \\
            0 & \gpp(\eta_1-\eta_2)
        \end{pmatrix}
        \begin{pmatrix}
            \vec{x}_2 \\ \vec{p}_2
        \end{pmatrix}
        ] \Theta(\eta_1-\eta_2).
    \]
    Indeed, it can be checked easily that its Fourier transform~\eqref{eq:free_evolution} is the inverse of~\eqref{eq:inverse_free_propagator}.
    
    \section{Proof of the closed-form kernels in one dimension}\label{app:proof_1D}
    In principle, the formalism presented in section~\ref{sec:perturbation_theory} is valid in an arbitrary number of spatial dimensions. For the one-dimensional case, however, things simplify considerably since inner products are replaced by ordinary associative products. In three spatial dimensions, structure formation along an axis perpendicular to sheets of homogeneous density can be thought of as an effectively one-dimensional system, leaving the background dynamics unchanged~\cite{McQuinn2016}.
    Given this setup and for the CICs specified in section~\ref{sec:perfectly_cold}, we will prove that the growing-mode Vlasov kernels of arbitrary order have the closed-form solution
    \[\label{eq:general_1d}
        \mathcal{F}_n(1; q_1,\dots,q_n) = 2\pi \dirac\qty(k_1+q_{1\dots n})\,\bar{\rho}^{1-n}\,\e^{n\eta_1}\,\frac{(q_{1\dots n} - l_1)^n}{n! \prod_{j=1}^n q_j},
    \]
    recovering known results~\cite{Grinstein1987,McQuinn2016} for the SPT/LPT density kernel, i.\,e.~for $l_1=0$.
    We start by considering the linear solution
    \[
    \mathcal{F}_1(1; q_1) = 2\pi \dirac\qty(k_1+q_1)\,\e^{\eta_1}\,\frac{q_1-l_1}{q_1},
    \]
    which indeed fulfils~\eqref{eq:general_1d}, serving as a starting point for a proof by induction. Suppose now that $\exists n\in\mathbb{N}_{>0}: \forall m\in\mathbb{N}_{>0}$ with $m<n$, \eqref{eq:general_1d} holds. Inserting it into the recursion relation~\eqref{eq:recursion_vlasov}, the innermost integrations yield
    \begin{multline}\label{eq:proof_innermost}
        \gamma(2,-3,-4)\frac{1}{n!}\sum_{\pi\in S_n}\sum_{m=1}^{n-1}\mathcal{F}_m(3;q_{\pi_1},\dots,q_{\pi_m})\,\mathcal{F}_{n-m}(4;q_{\pi_{m+1}},\dots,q_{\pi_n})\\
        = 2\pi\dirac(k_2+q_{1\dots n})\,\frac{3}{2} \bar{\rho}^{1-n}\,\e^{n\eta_2}
        \frac{-l_2}{n!\prod_{j=1}^n q_j} \sum_{\pi\in S_n}\sum_{m=1}^{n-1}\frac{q_{\pi_1\dots\pi_m}^{m-1}}{m!} \frac{\qty(q_{\pi_{m+1}\dots\pi_n}-l_2)^{n-m}}{(n-m)!}.
    \end{multline}
    Ignoring the symmetrization at first, we can use the binomial theorem in order to split off powers of $l_2$,
    \[\label{eq:innermost_split}
        \sum_{m=1}^{n-1}\frac{q_{1\dots m}^{m-1}}{m!} \frac{(q_{m+1\dots n}-l_2)^{n-m}}{(n-m)!}
            &= \frac{1}{n!} \sum_{m=1}^{n-1} \binom{n}{m} q_{1\dots m}^{m-1}\,(q_{m+1\dots n}-l_2)^{n-m} \\
            &= \frac{1}{n!} \sum_{m=1}^{n-1}\sum_{r=0}^{n-m} \binom{n}{m}\binom{n-m}{r} q_{1\dots m}^{m-1}\,q_{m+1\dots n}^{n-m-r} (-l_2)^r \\
            &= \frac{1}{n!} \qty[ A(q_1,\dots,q_n) + \sum_{r=1}^{n-1}\frac{(-l_2)^r}{r!} B_r(q_1,\dots,q_n) ],
    \]
    where in the last step we separated terms with $r=0$ and reordered the remaining sums, defining
    \[
        A(q_1,\dots,q_n) &\equiv \sum_{m=1}^{n-1} \binom{n}{m} 	q_{1\dots m}^{m-1}\,q_{m+1\dots n}^{n-m} \\
        B_r(q_1,\dots,q_n) &\equiv \sum_{m=1}^{n-r} \frac{n!}{m!(n-m-r)!} q_{1\dots m}^{m-1}\,q_{m+1\dots n}^{n-m-r}.
    \]
    Note that $A$ and $B_r$ are the only ingredients in the recursion relation~\eqref{eq:recursion_vlasov} which are not trivially symmetric under permutations of the $q_i$. By the multinomial theorem, we have
    \[
        \frac{1}{n!} \sum_{\pi\in S_n} B_r(q_{\pi_1},\dots,q_{\pi_n})
        &= \sum_{m=1}^{n-r} \frac{1}{m} \sum_{\substack{\beta_1,\dots,\beta_m \geq 0\\\beta_1+\dots+\beta_m=m-1}} \sum_{\substack{\beta_{m+1},\dots,\beta_n \geq 0\\\beta_{m+1}+\dots+\beta_n=n-m-r}} \sum_{\pi\in S_n} \prod_{i=1}^n \frac{q_{\pi_i}^{\beta_i}}{\beta_i!} \\
        &= \sum_{\substack{\beta_1,\dots,\beta_n \geq 0\\\beta_1+\dots+\beta_n=n-1-r}} \prod_{i=1}^n \frac{q_i^{\beta_i}}{\beta_i!} \sum_{m=1}^{n-r} \frac{1}{m} \sum_{\pi\in S_n}[\beta_{\pi_1}+\dots+\beta_{\pi_m}=m-1],
    \]
    with the Iverson bracket $[\cdot]$ equating to unity if the condition is fulfilled and zero otherwise.
    For fixed values $\beta_i$, let
    \[
        \mathfrak{I}_m \coloneq \{(i_1,\dots,i_m)\in [1,n]^m \mid (\forall a,b\in[1,m]: i_a\neq i_b) \wedge (\beta_{i_1}+\dots+\beta_{i_m} = m-1)\}
    \]
    denote the set of ordered index tuples of length $m$ fulfilling the sum condition.
    Then the innermost sum of the last line is
    \[
        \sum_{\pi\in S_n}[\beta_{\pi_1}+\dots+\beta_{\pi_m}=m-1] = (n-m)!\,\abs{\mathfrak{I}_m},
    \]
    since every non-vanishing term belongs to a permutation that can be decomposed into an element of $\mathfrak{I}_m$ and some permutation of the remaining $n-m$ indices. Let now $\mathfrak{C}_m$ denote the cycles of length $m$, i.\,e.~the set of equivalence classes in $\mathfrak{I}_m$ under cyclic permutations. Since there are $m$ cyclic permutations of $m$ elements, we have
    \[\label{eq:permutation_sum}
        \sum_{m=1}^{n-r} \frac{1}{m} \sum_{\pi\in S_n}[\beta_{\pi_1}+\dots+\beta_{\pi_m}=m-1] = \sum_{m=1}^{n-r} (n-m)!\,\frac{\abs{\mathfrak{I}_m}}{m} = \sum_{m=1}^{n-r} (n-m)!\,\abs{\mathfrak{C}_m}.
    \]
    This sum counts the number of permutations $\pi\in S_n$ that can be decomposed into some cyclically ordered block $C\in\bigcup_{m=1}^{n-r} \mathfrak{C}_m$ and a linearly ordered ``tail'' tuple $T$ of length~$n-\abs{C}$.
    
    Now take some permutation $\pi\in S_n$. We want to show that we can always identify it with a tuple $(C,T)$ as above. In order to do so, find the smallest non-negative $m$ such that $\beta_{\pi_1}+\dots+\beta_{\pi_m} = m-1$, i.\,e.~$C=(\pi_1,\dots,\pi_m) \in \mathfrak{C}_m$. Such an $m$ always exists: Without loss of generality, we have $\beta_{\pi_1}> 0$, since otherwise we would be done by setting $C=\{\pi_1\}$. Since all $\beta_i$ are non-negative, the partial sums $\Sigma_k = \sum_{i=1}^k \beta_i$ are monotonically increasing with $k$. However, $\Sigma_n = n-1-r < n$. Therefore, $\exists m: \Sigma_{m-1}>m-2$ and $\Sigma_{m+1}<m$ and thus $m-2<\Sigma_m<m \implies \Sigma_m = m-1$. We know that $m\leq n-r$, because otherwise $\Sigma_n>n-r-1$ by monotonicity. Hence, we can split $\pi$ at this minimal $m$ into a cyclically ordered block $C$ and a linearly ordered tail $T$. This defines an injection from $S_n$ into the set of such tuples $(C,T)$. In fact, this map is a bijection.
    This can be seen from the Dvoretzky-Motzkin cycle lemma~\cite{Dvoretzky1947}, which in a special case can be stated as~\cite{Dershowitz1990}
    \begin{quote}\textsc{Cycle Lemma} (special case).\quad
        For any sequence of $p_1 p_2\dots p_{2m-1}$ of $m$ ``boxes'' and $m-1$ ``circles'', there exist exactly one cyclic permutation~$p_j p_{j+1}\dots p_{2m-1}p_1\dots p_{j-1}$, $1\leq j\leq 2m-1$, that is dominating.
    \end{quote}
    A sequence $p_1 p_2 \dots p_l$ of boxes and circles is called \emph{dominating} if for any $i$, $1\leq i\leq l$, the number of boxes in $p_1 p_2\dots p_i$ is larger than the number of circles.
    
    We can think of the boxes as boundaries of (circularly arranged) bins and the circles between them as balls in the bins. The number of balls in the $i$-th bin (of a total of $m$ bins) is then identified with the value of $\beta_i$. Given some $C\in\bigcup_{m=1}^{n-r} \mathfrak{C}_m$, the cycle lemma tells us that we can find one special rotation such that $m$ is the minimal index fulfilling the sum property. We construct a permutation $\pi\in S_n$ from a tuple $(C,T)$ as above by identifying the first $m$ elements of $\pi$ with this special rotation of $C$. The remaining $n-m$ elements are fixed by $T$. Hence, every tuple $(C,T)$ is mapped to by the previously defined map exactly once, making it a bijection. As a result, the sum~\eqref{eq:permutation_sum} equals $\abs{S_n}=n!$ and
    \[
        \frac{1}{n!} \sum_{\pi\in S_n} B_r(q_{\pi_1},\dots,q_{\pi_n}) = \frac{n!}{(n-1-r)!}\, q_{1\dots n}^{n-1-r}
    \]
    by the multinomial theorem. Similarly,\footnote{$A$ can be obtained from $B_r$ by setting $r=0$ and subtracting the term $m=n$.} we obtain
    \[
        \frac{1}{n!} \sum_{\pi\in S_n} A(q_{\pi_1},\dots,q_{\pi_n}) = (n-1)\,q_{1\dots n}^{n-1}.
    \]
    Inserting these into~\eqref{eq:innermost_split} leads from~\eqref{eq:proof_innermost} to
    \begin{multline}
        \gamma(2,-3,-4)\frac{1}{n!}\sum_{\pi\in S_n}\sum_{m=1}^{n-1}\mathcal{F}_m(3;q_{\pi_1},\dots,q_{\pi_m})\,\mathcal{F}_{n-m}(4;q_{\pi_{m+1}},\dots,q_{\pi_n})\\
        = 2\pi\dirac(k_2+q_{1\dots n})\,\frac{3}{2} \bar{\rho}^{1-n}\,\e^{n\eta_2}
        \frac{-l_2}{n!\prod_{j=1}^n q_j} \qty(n(q_{1\dots n}-l_2)^{n-1} - q_{1\dots n}^{n-1}).
    \end{multline}
    Next, we apply the free evolution~\eqref{eq:free_evolution}, replacing
    \[
        k_2 \to -q_{1\dots n} \qq{and} -l_2 \to \gqp(\eta_1-\eta_2)\,q_{1\dots n} - \gpp(\eta_1-\eta_2)\,l_1.
    \]
    Collecting powers of $l_2$ by means of the binomial theorem, the remaining time integrations are of the form
    \begin{multline}
        \int_{0}^{\eta_1}\dd\eta_2\,\e^{n\eta_2} (\gqp(\eta_1-\eta_2)q_{1\dots n} - \gpp(\eta_1-\eta_2)l_1)^r %&= \sum_{s=0}^{r}\binom{r}{s} q_{1\dots n}^{r-s} (-l_1)^s \int_{0}^{\eta_1}\dd\eta_2\,\e^{n\eta_2} \gqp(\eta_1,\eta_2)^{r-s}\,\gpp(\eta_1,\eta_2)^s \\
        \\\approx \e^{n\eta_1}\sum_{s=0}^{r}\frac{r!}{s!}\frac{(2n+s-1)!}{(2n+r)!} 2^{r-s+1} q_{1\dots n}^{r-s} (-l_1)^s,
    \end{multline}
    where we only kept the fastest-growing mode by setting the lower integration boundary to~$-\infty$ and used the defining identity of the beta function\footnote{$\mathrm{B}(z_1,z_2) = \int_{0}^{1}\dd u\,u^{z_1-1}(1-u)^{z_2-1} = \frac{\Gamma(z_1)\Gamma(z_2)}{\Gamma(z_1+z_2)}\quad\forall z_1,z_2\in\mathbb{C},\ \Re(z_1),\Re(z_2)>0$} to solve the integral. Upon some rearranging, the sums over $r$ and $s$ can be evaluated to obtain
    \begin{multline}
        g_0(1,-2)\gamma(2,-3,-4)\frac{1}{n!}\sum_{\pi\in S_n}\sum_{m=1}^{n-1}\mathcal{F}_m(3;q_{\pi_1},\dots,q_{\pi_m})\,\mathcal{F}_{n-m}(4;q_{\pi_{m+1}},\dots,q_{\pi_n})\\
        %= 2\pi\dirac(k_1+q_{1\dots n})\,\frac{3}{2}\bar{\rho}^{1-n}\,
        %\frac{\e^{n\eta_1}}{n!\prod_{j=1}^n q_j} \Biggl[\sum_{r=1}^{n}r\binom{n}{r} q^{n-r} \sum_{s=0}^{r}\binom{r}{s}(2q)^{r-s}(-2q-l_1)^s \frac{1}{n+\frac{s}{2}}\\
        %-\qty(\frac{2q}{n}-\frac{2q+l_1}{n+\frac{1}{2}})q^{n-1}\Biggr]\\
        = 2\pi\dirac(k_1+q_{1\dots n})\,\bar{\rho}^{1-n}
        \frac{\e^{n\eta_1}}{n!\prod_{j=1}^n q_j} \qty[(q_{1\dots n}-l_1)^n + 3 \qty(\frac{1}{n+\frac{1}{2}} - \frac{1}{n}) q_{1\dots n}^n + \frac{3\,l_1}{2n+1} q_{1\dots n}^{n-1}].
    \end{multline}
    Note that this is precisely the desired expression~\eqref{eq:general_1d} up to two additional terms. The final step is the application of $ [\identity+\omega]$, where the $\omega$-contribution exactly cancels the aforementioned terms in the fastest growing mode, concluding the proof of the closed-form solution~\eqref{eq:general_1d} $\forall n \in \mathbb{N}_{>0}$.

    \bibliographystyle{JHEP}
    \bibliography{references_bibtex}

\end{document}